\documentclass[preprint,12pt,authoryear]{elsarticle}

\usepackage{amssymb}
\usepackage{amsthm}
\usepackage{graphicx}
\usepackage{epsfig}
\usepackage{epstopdf}
\usepackage{bibentry}  
\usepackage{pdflscape}

\begin{document}
\journal{}
%\journal{Journal of Banking and Financing}
%\date{October 12, 2015}

\begin{frontmatter}
\title{Positive skewness, anti-leverage, reverse volatility asymmetry, and short sale constraints: Evidence from the Chinese markets}
%\tnoteref{t1}
%\tnotetext[t1]{This document is a collaborative effort.}
\author{Liang Wu} %, Jingyi Luo, Yingkai Tang, Greg} %\fnref{fn1}}
%\ead{liangwu@scu.edu.cn}
%\address{The School of Economics, Sichuan University,
%Chengdu,China 610065}
%{\cortext[cor1]{Corresponding author}}

\author{Jingyi Luo} %\fnref{fn2}}
%\ead{luojingyi@stu.scu.edu.cn}
\address{The School of Economics, Sichuan University,
Chengdu,China 610065}
%\fntext[fn1]{This is the first author footnote.}

\author{Yingkai Tang} %\footnote{Yingkai Tang is the corresponding author}}
%\ead{tang@scu.edu.cn}

\author{Gregory Bardes}
\address{The School of Business, Sichuan University,
Chengdu,China 610065}

\begin{abstract}
%It is interesting to note while a large literature documents the existing of positive return skewness and reverse volatility asymmetry of individual stocks in Chinese stock market. 
%There are some anomalies in the Chinese stock market, i.e., positive return skewness, anti-leverage effect, and reverse volatility asymmetry (positive return-volatility correlation). %There have been few papers aiming to explain the reasons backed by solid empirical evidence. 

There are some statistical anomalies in the Chinese stock market, i.e., positive return skewness, anti-leverage effect (positive returns induce higher volatility than 
negative returns); and reverse volatility asymmetry (contemporaneous return-volatility correlation is positive). 
In this paper, we first confirm the existence of these anomalies using daily firm-level stock return data on the raw returns, excess returns and normalized excess returns.
%, anti-leverage effect (positive leverage coefficient in the EGARCH model) and reverse volatility asymmetry (positive contemperaneous return-volaility correlations) for both raw and excess returns. 
We empirically show that the asymmetry response of investors to news is one cause of the statistical anomalies if short sales are constrained. Then in the context of slow adoption of security lending policy, we conduct panel analysis and empirically verify that the lifting of short sale constraints leads to significantly less skewness, less anti-leverage effect and less reverse volatility asymmetry. 
Positive skewness is a feature of lottery. 
%It has been shown by some researchers that there exists obvious lottery preference in the Chinese markets. 
Investors are encouraged to bet on the upside lottery like potentials 
%the continuation of price rising 
in the Chinese markets where the stocks skew more to 
the upside when short sales are constrained.
%Our research sheds a light on a possible cause of this phenomenon. 

%At last, our result also implies the mechanism through which short sale constraint affect return distribution and contemporaneous risk-return correlation, in hope that more thorough investigation be put into this issue

\end{abstract}


\end{frontmatter}

%×Ô¶¨ÒåÂÞÂíÊý×Ö
\makeatletter
\newcommand{\rmnum}[1]{\romannumeral #1}
\newcommand{\Rmnum}[1]{\expandafter\@slowromancap\romannumeral #1@}
\makeatother

%% main text
\section{Introduction}

Risk and return are fundamental concepts in finance. For example, the capital asset pricing model (CAPM) provides for an explicit relationship between risk and return by saying that the expected return of an asset is positively related to its undiversified risk. Accordingly, investors taking on additional risks should be compensated with higher expected returns. If the variance and covariance are time-varying, the dynamic nature between risk and volatility can be complex. Theories such as leverage effect (\cite{black1976stuedies,christie1982stochastic}) and volatility-feeback theory (\cite{pindyck1983risk,campbell1992no}) are developed to explain the asymmetric volatility property of individual stock returns in the United States (U.S.) which says that the stock return volatility is negatively correlated with stock returns (see e.g., \cite{bekaert2000asymmetric}).

The leverage effect hypothesis suggests that a drop in the value of a stock increases the debt-to-equity ratio, this 
in turn makes the stock riskier and causes future volatility to rise. 
Volatility-feedback theory suggests that a volatility increase raises the expected return, leading to an immediate stock 
price decline after investors revalue the equity. The causalities of the two theories are different: 
the leverage effect suggests that return shocks lead to the change of future volatility, while volatility-feedback 
suggests that current or future returns are caused by current or lagged volatility shocks. Since the two effects could 
happen at intraday sub-timescales, the observation of contemporaneous return-volatility correlation at a daily level 
could thus be caused by either effect or by both effects regardless of any differences in order of occurrence of return shocks and volatility shocks. 

Negative correlation of return and volatility has been observed in most of the financial markets in developed countries (\cite{bollerslev2008risk} (U.S. stock markets), \cite{qiu2006return} (German stock markets), \cite{bekaert2000asymmetric} (Japanese stock markets) and \cite{lee2012leverage} (Korea stock markets)). Other researchers such as \cite{harrison1999investigation,ghysels2005there,ludvigson2007empirical} find a positive relation. \cite{nelson1991conditional} and \cite{glosten1993relation} argue that either a positive or a negative relationship is possible. 

%%%%%%%%%%%%%%%%%%%%%%%%%%%%This is where I started to add data and method description in addition to the research results.%%%%%%%%%%%%%%%%%%%%%%%%%%%%%%%%%
While some researchers are discussing whether the correlation of return and volatility is negative or which theory seems to offer the best 
explanations, other studies focusing upon the Chinese markets reveals the existence of positive return-volatility 
correlation. Based upon daily raw returns from four market indices of the Shanghai and Shenzhen markets (taking from May, 1992 to August, 1996 for two composite indices and from October, 1992 to August, 1996 for two B-share indices), \cite{yeh2001interaction} find that good news has a larger impact upon future volatility than bad news. 
\cite{menggen2007risk} examines the log returns of both Shanghai and Shenzhen Composite Index. 
%from Dec. 19 1990 to Mar. 30 2007 for Shanghai Composite Price Index and from Apr. 3 1991 to Mar. 30 2007 for Shenzhen Component Price Index. 
He finds a positive and statistically significant risk-return relationship for the daily returns of stock indices composed of stocks traded on the Shenzhen Stock Exchange while the relationship is negative but insignificant for those referencing the Shanghai Stock Exchange. \cite{qiu2006return} discover the positive return-volatility correlation in Chinese stock indices using both daily and 5-min data of Chinese stock indices, compared with the negative correlation founded on German DAX Index. They refer to this as the anti-leverage effect, and their result is echoed by \cite{shen2009return}, who use a retarded volatility model and further argue that leverage effect and the anti-leverage effect are independent of the probability distribution of returns and long-range time correlation of volatilities.  \cite{bo2014mini} review this effect together with other features of the Chinese stock markets including the volatility auto-correlation, the return-volatility correlation and the spatial structure at the firm level. \cite{wan2014reverse} confirm the positive correlation using 5-minute data of 10 individual stocks with high capitalization and high liquidity selected from different industries in the Hushen 300 Index and finds a positive return-volatility correlation for most of the individual stocks in the sample. 
%They also point out that price jumps play an important role in the significance of this positive correlation. For most of the individual stocks in the sample, the positive correlation is insignificant until jumps are totally eliminated in both return and volatility.

A concept closely related to the dynamic risk-return relationship is the skewness of returns. 
If the return-volatility correlation is larger than zero, one would expect to observe larger magnitudes of positive returns than negative 
returns and therefore expect more chances for obtaining extreme positive returns than extreme negative returns. 
This would lead to a positively skewed return distribution. In line with the volatility asymmetry, it is a familiar and 
stylized fact that stock returns, especially those of large firms, have negative skewness in developed 
countries (see \cite{campbell1992no,engle1993common,kim2004more}. However, numerous papers document the existence of positive return skewness 
of individual stocks in China. \cite{su1998risk} collect daily and weekly return data of market indices and find that the skewness parameters 
of the distribution of both risk-adjusted returns and conditional standardized innovations are significantly positive for 
both A-shares and B-shares in Shanghai and ShenZhen. \cite{lee2001stock} study a value-weighted average market capitalization 
index as well as an equally weighted index of Shanghai and Shenzhen A-shares and B-shares based on the daily returns from 
December, 1990 to December, 1997 and also finds the existence of positive skewness. \cite{gu2008empirical} explore the ultra-high-frequency data of 23 stocks traded in the Chinese stock markets in 2003 and achieve similar results.

Possible explanations of the anomalies in the Chinese markets given by previous researchers are based on indigenous trading mechanisms. 
\cite{wan2014reverse} postulate that the Chinese markets protect investors against daily loss by limiting a stock's daily range to 
$\pm 10\%$ of the previous day's closing price, which reduces the risk averse of investors. 
Few or even no dividends cut down their motivations for value investment. The majority of the participants are retail investors who 
behave like short-term speculators. They surge for a price rising trend. \cite{klossner2012modeling} develop a statistical approach to detect intraday overreaction.
They find that Chinese investors
overeact to good news with greater emphasis than to bad news, when they are compared with investors in the U.S. and German markets who overeact
with greater emphasis to bad news than to good news.

However, among all of the empirical studies, few notice the linkage between short sale constraints and the anomalies observed in Chinese markets. Two contradicting theories were previously introduced to examine the effects imposed upon return skewness by short sale constraints. In an effort to explain market crashes, \cite{hong2003differences} show that based upon the difference among investor opinions and in the presence of short sale constraints, more information and thus 
a larger scale of price adjustment comes out when the market is falling, which results in negative skewness of a single traded asset, either for an 
individual stock or for the market portfolio. Another model developed by \cite{xu2007price} predicts that the equilibrium price is a convex 
function of information when short sale activity is constrained. He suggests that the market reacts more to a positive signal 
than to a negative signal, and that a convex transformation skews the market return to the right. Therefore, market returns should be more positively skewed.

Emprical studies seem to be in favor of \cite{xu2007price}'s  model. \cite{goetzmann2003efficiency} study weekly returns of value-weighted market indices from forty seven equity markets in the world and find that short sale constraints are associated with less negative, not more negative, skewness at the market level. In addition, they find that a severe difference of opinion, proxied by trading volume, would result in more skewness. \cite{saffi2010price} calculate skewness of weekly raw and excess returns of more than 12,600 stocks from 26 countries. They use firm-level lending supply as an indirect proxy for short sale constraints. 
After conducting panel regression, they find positive linkage between skewness and short sale constraints. They also argue that the relaxing of short sale constraints corrects overpricing and reduces the frequency and magnitude of positive extreme returns and thus leads to lower skewness. Supporting \cite{saffi2010price}, \cite{chang2007short} show that when short sale constraints are removed, both raw returns and excess returns of individual stocks exhibit less positive skewness, based on individual daily stock return data from Hong Kong. Exploiting the daily returns of ten of the fifteen off-limits stocks traded on the Paris stock market, \cite{lioui2010impact} points out that the short sale ban increases the uncertainty in the economy and is responsible for the deterioration of return asymmetry (i.e., the return skewness was exacerbated).
One exception is discovered by \cite{lin2010short} who study the daily returns of 33 constituent stocks of the Taiwan 50 Index Fund and find no significant difference in return skewness before and after stock short sale constraints are lifted. 

It is worth noting that it is difficult to directly examine the effect of short sales on return-volatility correlation and skewness, since short sales have long been practiced in developed markets. To this end, many researchers use proxies such as relative short interest, breadth of ownership, institutional ownership for the level of short sale constraints (see \cite{figlewski1981informational,chen2002breadth,nagel2005short}). The 2008 financial crisis, during which short sales were banned to different extents among different countries, provides researchers with the opportunity to test the effects of short sale constraints on pricing efficiency and market stability. Nevertheless, the credence and stability of their results are dampened by the brief periods of the bans and the limited number of affected stocks (e.g. the short ban lasted for only 3 weeks in the U.S.). In this respect, the Chinese stock market is ideal for empirical testing: the launches of margin financing and security lending for selected stocks 
%after March 31st 2010 
naturally divides the time horizon into two time periods (i.e., before and after the stocks can be sold short). The fact that stocks can be added to the short sale list only if they meet certain requirements provides us with a test group as well 
as a control group composed of shortable and non-shortable stocks, respectively. 

Based on the margin financing and security lending policy, two previous papers (\cite{chang2014short},\cite{zhao2014short}) studying the effect of short sales focus mainly on market efficiency. The authors find that after the ban is lifted, price efficiency increases and stock return volatility decreases.  
In addition, the authors have reported the change of skewness of raw returns around the addition events (eligible stocks are added to a designated list for margin-trading and short-selling). However, their results are in conflict with each other. \cite{chang2014short} find higher skewness when short sales are allowed by conducting an event study of 285 additions between March 31, 2010 and November 25, 2011. Similarly, \cite{zhao2014short} apply an event study to 511 addition events, between February, 2010 and August, 2013, they observe lower skewness after short sale constraints are lifted.

The first motive of this paper is to test whether skewness of return distribution and return-volatility correlation are positive for daily firm-level returns in the Chinese markets. Most previous studies of the Chinese markets focus on stock indices (see \cite{yeh2001interaction,menggen2007risk,qiu2006return} for return-volatility correlation and \cite{su1998risk,lee2001stock,gu2008empirical} for skewness). One contribution of our study to the literature is that we have a thorough test of the statistics based on raw returns, excess returns and normalized returns of individual stocks. The excess returns 
retain the idiosyncratic aspects of each stock while the market aspect is removed. The normalized excess returns, on the other hand, remove the effect of time-varying volatility and possible leverage (or anti-leverage) effect through dividing the excess returns 
by the conditional volatility obtained from the EGARCH model.

%While most previous studies of the Chinese markets focus on stock indices (see \cite{yeh2001interaction,menggen2007risk,qiu2006return} for return-volatility correlation and \cite{su1998risk,lee2001stock,gu2008empirical} for skewness). 

%As our result shows, positive skewness are obtained for the raw returns, excess returns and normalized excess returns. The skewness of excess returns are the most significantly larger than 0 after the market returns are removed from individual stock returns. The normalized excess returns are slightly smaller after the effect of anti-leverage effect is removed. But positive skewness still remains. Then we check the coefficient of the leverage effect in EGARCH model as well as contemporaneous return-volatility correlation coefficient for both raw and excess returns. In line with the findings by \cite{wan2014reverse} that the positive correlation between volatility and return not only occurs to lagged returns but also to contemporaneous ones, we make it clear that short sale lead to both positive contemporaneous risk-return correlation thus to positive skewness while they suggest people’s rushing for stock prices as one possible reason for this positive correlation.

The second contribution of our study is that we have empirically shown that short sale constraints are responsible for the existence of positive skewness, anti-leverage effect and reverse volatility asymmetry in the Chinese markets, based on a panel regression on a wide range of daily return data covering both shortable and non-shortable stocks spanning from 2006 to 2014 (4 years before and 4 years after the first launch of security lending policy). Some researchers believe that these statistical anomalies are caused by the fact that investors in the Chinese markets tend to be speculators and rush for a price rising (see e.g.\cite{qiu2006return,wan2014reverse}). It is unlikely that Chinese investors are significantly different from other investors throughout the world. There should be some mechanisms built into the markets which offer the Chinese market investors the motivation to behave differently. In this paper, we present evidence that one possible mechanism to explain the statistical anomalies resides within the policy of the short sale constraints. When the short sales are constrained, investors tend to rush for a price rising hoping to make profits since they can only buy low and sell high. As a result, we can observe that the market reacts more to a positive signal than to a negative signal, 
and hence the presence of positive skewness, anti-leverage effect, and reverse volatility asymmetry.

%Going further from confirming the existence of positive skewness and return-volatility correlation, the casual effect of short sale on skewness and return-volatility correlation is barely examined, and a large literature document the influence of short sale on stock market and they are mainly centered on information transmission and price efficiency. For example see....Therefore, the second objective of this paper is to test the proposition that the lift short sale constraint results in less positive skewness and risk return correlation. Different with \cite{chang2014short} and \cite{zhao2014short}, we first present descriptive data of return asymmetry in our own measurement and positive skewness of individual stocks in China. And then based on a wide range of daily return data covering both shortable and non-shortable stocks from 2006 to 2014, we conduct a panel regression in which five major adjustments of the designated list are denoted by dummy variables. Our results significantly support \cite{xu2007price}theoretical prediction that skewness should decrease with fewer short-sale constraints and run contradictory to \cite{chang2014short}'s empirical finding that higher skewness would appear when short sales are allowed. 

The remainder of this paper is organized as follows: part 2 gives a brief description of the data set; part 3 presents the data processing method and empirical results; part 4 is the conclusion and discussion.

\section{Data}
%\subsection{Short Sales in the Chinese markets}
We collect Chinese stock data and the designated short sale list from the GuoTaiAn database. We only keep data on A-share stocks traded on the Chinese main board. We exclude the stocks traded in the China Growth Enterprise Market which begin its trading activities on October 23, 2009, significantly after the moment of our earliest data selection. Turnover rate data and the FAMA three-factor data are obtained from the Wind Investor Terminal (WIT). 

Short sale practices started on March 31, 2010 in China. Only a list of designated securities that meet certain requirements can be sold short. Up to March 31, 2014, there are 756 securities allowed for short sale through the margin trading and security lending policy, including 742 stocks and 14 ETFs listed on the Shanghai and Shenzhen Stock Exchanges. There are a total of 2,674 A-share stocks traded on the Chinese main board up to March 31, 2014. Among them, 742 stocks are on the designated list, and 1,932 stocks are not. Table \ref{tab:events} reports the timeline of the gradual lift of the short sale ban in China. Similar to \cite{chang2014short}, we organize the table according to the events in which stocks are added to or deleted from the designated list. We have a longer list than reported by \cite{chang2014short} since more events are now available. As we can see from the table, the adoption of the security lending policy takes a rather long time, which gives us enough time and data to test the effects of the policy.

There are 2,674 A-share stocks in total in our samples as of March 31, 2014. Starting from January 1, 2002, four years of data are rolled to estimate the FAMA three-factor model to obtain excess returns. Thus the actual effective period is January, 2006 to March, 2012, eight years in total; four years before the launch of margin financing and security lending, and four years after that. Only those stocks with more than sixty days of data are kept after the calculation of excess returns. Thus, we end up with 1,357 stocks in our samples (640 are on the designated list and 717 are not). For comparison, we choose 3,023 stocks in the Russel-3000 index, traded in the U.S. markets for comparison. After the excess returns are calculated, we end up with 2,552 stocks in our samples for the U.S. Markets. The FAMA three factor data is obtained from French's website\footnote{http://mba.tuck.dartmouth.edu/pages/faculty/ken.french/data \_library.html}.

 %Note that the list of shortable stocks are revised for a few times, as shown in Table 1. 

%As a comparative group, we exclude 742 target stocks from the total list of 2788 stocks and excluded 114 B shares and thus get 1932 stocks that cannot be shorted. Since the total list is obtained in January 2015 while some stocks are listed only after March, 2014 and some are delisted before 2002, 1849 of the 1932 nontarget stocks have trading data during the 12 years from January 2002 to March 2014. These stocks that cannot be shorted are referred to as nontarget stocks, non-shortable stocks or Group 2 in this paper.

%The number of objects changes as we conduct further calculation. The calculation of skewness are based on the distribution of raw return, abnormal return and normalized return of stocks for every six-month. If a stock does not have enough trading data for 44 days (approximately two month) for a certain six-month period, then its return skewness for this period is not calculated. And the number of target stocks are reduced to 640 and nontarget stocks to 718 after calculating Ratio This is because we require return data of at least a continuous 10 days for computing Ratio, and stocks with Ratio larger than 20, indicating an unreasonably strong asymmetry in excess return, are dejected. So the ultimate effective number of target and nontarget stocks used in panel data regression, are 640 and 718 respectively.  

\begin{table}
    \caption{The table reports the statistics on the events in which stocks in the Chinese markets are allowed for short sale. Effective date is the day on which the short sale can actually be practiced for those stocks on the list.}
\label{tab:events}
\begin{tabular}{lccccc}
    \hline
       & Effective date & No. added & No. deleted  & No. on list \\
    \hline
    	 & 2010/03/31     & 90        & -             & 90 \\
   		 & 2010/07/01     & 5         & 5             & 90 \\
   		 & 2010/07/29     & 1         & 1             & 90 \\
   		 & 2011/12/05     & 189       & 1             & 278 \\
   		 & 2012/06/04     & 2         & -             & 280\\
   		 & 2012/10/29     & 1         & -             & 281\\
   		 & 2013/01/31     & 276       & -             & 557\\
   		 & 2013/03/06     & -         & 1             & 556\\
   		 & 2013/03/07     & -         & 1             & 555\\
   		 & 2013/03/26     & -         & 2             & 553\\
   		 & 2013/03/29     & 1         & 2             & 552\\
   		 & 2013/04/10     & 1         & -             & 553\\
   		 & 2013/04/24     & 1         & -             & 554\\
   		 & 2013/05/02     & -         & 1             & 553\\
   		 & 2013/05/03     & -         & 1             & 552\\
   		 & 2013/05/27     & 1         & -             & 553\\
   		 & 2013/07/25     & 1         & -             & 554\\
   		 & 2013/08/05     & -         & 2             & 552\\
   		 & 2013/09/16     & 205       & -             & 757\\
   		 & 2013/12/04     & 1         & -             & 758\\
   		 &Cumulated       &775        &17             &758 \\
    \hline
  \end{tabular}
\end{table}

\section{Empirical study}
In this section, we want to show empirically that positive skewness of return distribution, anti-leverage effect, and the reverse volatility asymmetry (positive return and risk correlation) in the Chinese financial markets are all caused at least partially by short sale constraints. First, we present the data processing method.

\subsection{Data processing}
In order to confirm the existence and the robustness of positive skewness as well as positive risk-return correlation in the Chinese markets, we conduct our experiment on the raw stock returns, excess returns, and normalized excess returns.  

%And then we demonstrate the existence of asymmetry in excess return, i.e., we point out the difference between the magnitude of positive excess return and its negative counterpart. Further more, we exemplify the possible influence of short sale policy by making a comparison between target stocks and non-target ones. In the end we conduct a categorical regression in which both the passage of time and the status of a stock--whether it can be shorted or not, are set as dummy variables, and the result of which again confirms our suggestion.%ÕâÀïÒª²»ÒªÌá½áÂÛ£¿

Excess return is obtained from the FAMA three-factor model for each individual stock.
\begin{equation}
\label{eq:excess}
\epsilon_{t} = y_{t} - (\alpha+\beta_{1}r_{t1}+\beta_{2}r_{t2}+\beta_{3}r_{t3})
\end{equation}
where $\epsilon_{t}$ is the excess return (or innovation), $r_1, r_2$ and $r_3$ respectively, represent one of the FAMA three factors, namely market return, market capitalization, and price-to-book ratio. The parameters $\alpha$ and $\beta$s are estimated from a four-year rolling time window and the excess return is computed for the next one quarter. In this way, the parameters are gradually adapted and do not cause large disruption to $\epsilon_{t}$.

We then normalize the excess return by dividing its conditional standard deviation obtained from the EGARCH model.
 %And for further observation on information delivery efficiency, not only daily returns, but return of different return intervals of every 2, 3, 4...10 days are also calculated and regressed under three-factor model.
We assume that $\epsilon_{t}$ is sampled from a normal distribution $\mathcal{N}(0,\sigma_t^2)$. For the evolution equation of $\sigma_t^2$, as has been documented by many researchers to be suitable for the financial time series in the Chinese markets, the EGARCH (1, 1) model is deployed here to estimate the variances of residuals.
\begin{eqnarray}
\label{eq:egarch}
&\log\sigma^{2}_{t}=\kappa+\gamma_{1}\log\sigma^{2}_{t-1}+\eta_{1}[\frac{|\epsilon_{t-1}|}{\sigma_{t-1}}-\sqrt\frac{2}{\pi}]
+\xi_{1}(\frac{\epsilon_{t-1}}{\sigma_{t-1}}) \\
\label{eq:norm}
&\epsilon'_{t} = \epsilon_{t}/\sigma_t 
\end{eqnarray}
where $\epsilon'_{t}$ is the normalized excess return.

The excess return $\epsilon_t$ keeps the idiosyncratic innovations of each stock while the market part is removed. The normalization of excess return removes the effect of time-varying  volatility by dividing the conditional volatility estimated from the EGARCH model. The other advantage of the normalization is that the leverage effect has also been removed after the excess return is normalized by the EGARCH model, which has a leverage term. By the assumption of the EGARCH model, we know that even if a leverage or an anti-leverage effect is present, $\epsilon_t$ is still symmetrically distributed if it is sampled from $\mathcal{N}(0,\sigma_t^2)$. Therefore, if we can still observe asymmetry in the normalized excess returns, the asymmetry is not due to the leverage or anti-leverage effect described in the EGARCH model.

\subsection{Statistics of Overall Markets}
In this section, we present some stylized statistical facts in the Chinese markets in comparison with the U.S. markets. 

\begin{figure}
\centering 
\includegraphics[width =0.7\textwidth]{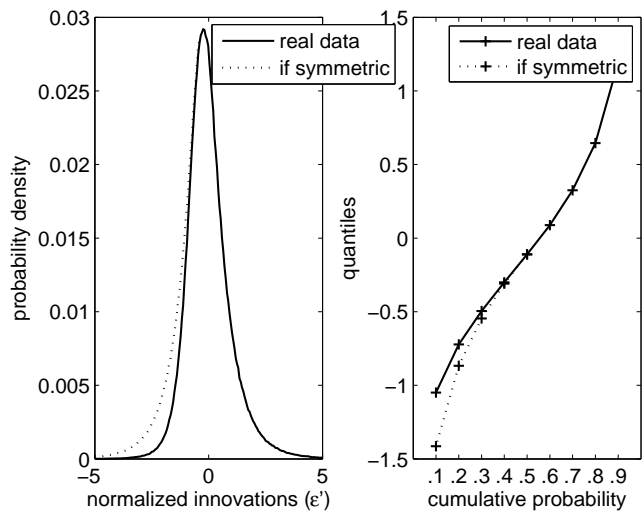}
\caption{The probability density and distribution function of normalized excess returns. The left side graph is the probability density function and the right side graph is the distribution function. The dotted line in the right graph is the reflection of the distribution function of the right tail, which is heavier than the left tail.}
\label{fig:return}
\end{figure}

First, we demonstrate graphically that the distribution of daily firm-level returns in the Chinese markets is positively skewed. The daily normalized excess returns are accumulated for all stocks (regardless of whether the stocks are on the designated list for short sales). The probability distribution and the distribution function are plotted in Figure \ref{fig:return}. We can clearly see from the Figure that the right tail of the distribution is heavier and therefore the returns are positively skewed.

We then numerically calculate skewness, leverage coefficent $\xi_1$ and return-volatility correlation for stocks traded in both the Chinese and the U.S. markets. Skewness reported in this study is normalized skewness (i.e., the third centered moment normalized by the standard deviation cubed). All of the statistics are calculated first on daily returns for each stock, then the mean and median are estimated to test whether the statistics are larger or smaller than zero. Skewness is calculated in three different ways: namely, skewness of the raw returns, skewness of the excess returns, and skewness of the normalized excess returns. The skewness of excess returns captures the asymmetry of excess returns of each stock after the market part is removed. The skewness of normalized excess return captures the asymmetry of excess returns after the time-varying volatility and leverage effects are removed.

As the first measurement of the risk-return relationship, the leverage coefficient $\xi_1$ in the EGARCH model is deployed as a proxy. The leverage coefficient captures the change of future volatility caused by past returns. Then we compute the daily contemporaneous return-volatility correlation directly (i.e., both return and volatility are employed for the same day). The absolute value of the daily return is used to proxy for the contemporaneous volatility. 

The results of the descriptive statistics are reported in Table \ref{tab:des}. In China, the mean and median of skewness are both positive at a 1\% significance level, which confirms the positive skewness we have observed in Figure \ref{fig:return}. Note that the skewness of the aggregate market returns is usually negative. For example, the skewness of the value weighted market returns is -0.32 (not reported in the Table \ref{tab:des}) for the same time period. After removing the negative part of the aggregate market, the skewness of the excess returns of individual stocks is much higher than that of the raw returns. The origin of the positive skewness comes from the idiosyncratic innovations of stocks. The skewness of excess returns and normalized excess returns are close in their values. The removal of time-varying volatility and leverage effect has little impact upon the skewness after the time-varying volatility and leverage effect are removed. As a sharp contrast, both mean and median of skewness are negative at a 1\% level in the U.S. markets. It has been reported in \cite{albuquerque2012skewness} that the average skewness is positive. The possible reasons for the difference are: (i) The average value of non-normalized skewness (not divided by the standard deviation cubed) is reported in their study. Small firms tend to be more volatile and  have returns with higher positive non-normalized skewness. Thus, small firms have dominating weights when an average is taken together with both large firms and small firms included; (ii) our research uses the top 3000 stocks as in the Ruessel index so that we have a comparable number of stocks and market capitalization for the two countries. Thus, we have a lesser number of small firms included in our study while they use all stocks in the CRSP (Center for Research in Security Prices) database. It should be noted that although we calculate the statistics for the whole time period, the results are similar if we divide the data into half-year time windows (not reported).

As for the risk-return relationship, positive mean and median leverage coefficients indicate that we have an anti-leverage effect in China in contrast with a leverage effect in the U.S. In addition, the contemporaneous return-volatility correlation is also positive in China. The results in the U.S. show no significant difference from zero.

\begin{landscape}
\begin{table}
\caption{The table reports the overall descriptive statistics of daily firm-level return in China and the U.S.. The skewness is measured for the raw return, excess return and normalized excess return for each stock. The leverage coefficient $\xi_1$ in the EGARCH model is deployed to measure the leverage (or anti-leverage) effect. Contemporaneous correlation of return and its absolute value is reported to measure the volatility asymmetry. A T-test and a signed rank test are, respectively, used to test the sign of mean and median. In China, we have positive values for skewness, leverage coefficient, and risk-return correlation. However, in the U.S., skewness and leverage coefficients are both negative.}
\label{tab:des}
\begin{tabular}{lcccccl}
%\hline 
%&\multicolumn{3}{c|}{skewness} &  & \multicolumn{2}{c}{correlation} \\
%& raw & excess & norm. & leverage & raw & excess \\
    \hline
    &\multicolumn{3}{c}{skewness} & leverage & \multicolumn{2}{c}{return-volatility correlation} \\
    \cline{2-4}
    \cline{6-7}
	& raw & excess & normalized & coefficient & raw & excess \\
\hline 
mean(C) & $0.69^{***}$ &   $1.36^{***}$  &  $1.24^{***}$ &   $0.02^{***}$ &   $0.08^{***}$ &   $0.28^{***}$ \\
mean(U) &$-0.23^{***}$  & $-0.27^{***}$  & $-0.23^{***}$ &  $-0.027^{***}$ &  $-0.0038$  &  $0.0052^*$ \\
median(C)& $0.075^{***}$ &  $ 0.94^{***}$ &   $1.02^{***}$  &  $0.038^{***}$ &   $0.058^{***}$  &  $0.30^{***}$ \\
median(U) &$-0.067^{***}$ &  $-0.026^{***}$  & $-0.01^{***}$ & $ -0.018^{***}$  &  0.0017 &   $0.01^{***}$ \\
\hline
\end{tabular}

\small{\emph{Note:Triple asterisks(***)denote significance at 1\% critical level, double asterisks(**) at 5\% level, and single asterisks(*) at 10\% level, respectively. C stands for China and U stands for the United States. Norm. is the abbreviation of normalized returns.}}
\end{table}
\end{landscape}

%It is mentioned in some work that individual stocks 

 %With no regarding to the time nor the group of stocks, we simply mix up the normalized innovations of all stocks from 2006 to 2014 and get the figure of distribution in the left part of Figure 1, from which the asymmetry can be clearly seen. A larger number of stocks are presenting positive innovations of a larger magnitude. In the right part of Figure 1, we calculate the cumulative probability of the distribution, again there's a perspicuous discrepancy between the real values and the presumed values where the excess return is symmetric.The departure of the distribution in reality from the distribution it should be if it is symmetric leads to the conclusion that the excess return is asymmetric.

\subsection{Effects of short sales in skewness and risk-return relationship}

Up to this point, we have observed that the Chinese markets are unique and they have quite different statistics. Some researchers believe that this is due to the fact that investors in the Chinese markets tend to be speculators and rush for a price rising (see e.g.\cite{qiu2006return,wan2014reverse}). We believe that it is unlikely that Chinese investors are significantly different from other investors throughout the world. There should be some mechanisms built into the markets which offer Chinese investors the motivation to behave differently. In this part, we present evidence that one possible mechanism to explain the statistical anomalies resides within the short sale constraints. 

The usual way to study the effect of short sale constraints is to take only a few dozen days around the events and thus the time effect and other market events may not be distracting issues. In our study, the time span is rather long (i.e., four years before the first launch of security lending policy and four years after). In order to exclude the possibility that positive skewness, anti-leverage effect and positive return-volatility correlation are caused by factors other than the short sale constraints, we include those stocks which are not yet eligible for short sales for comparison (thanks to the slow adjustment of the security lending policy). Therefore, we divide the stocks into two groups, namely, Group 1 which includes stocks that are selected for short sales practice at least once as of March 31, 2014, and Group 2 which includes stocks that are never eligible for short sales.

From Figure 2, we see a significant decrease in the skewness of returns for stocks in Group 1 after short sales are practiced, and the difference is most significant for daily returns. Looking at the results of Group 1 alone, the drop in the skewness may be caused by other reasons or by mere time effect. In contrast, the two intertwined lines at the top show that the skewness of stocks in Group 2 does not change much after the security lending policy is adopted. This result leads us to believe that short sale constraints should be responsible for return asymmetry, since skewness remains the same in the presence of short sale constraints for Group 2, while return asymmetry is largely ameliorated after the constraints are lifted. In addition, for both Group 1 and Group 2, skewness decreases gradually as the return interval increases, meaning that the deviance of positive and negative normalized excess returns is getting closer. The asymmetry of the price response to good news and to bad news starts to weaken after a couple of days but still persists even after 2 weeks (10 trading days). The persistence of positive skewness in the Chinese markets may be caused by the inadequate scale of short sale (e.g., on March 31, 2014, the outstanding short interest is only 5 billion RMB while the outstanding margin finance is 700 billion RMB) or other structural reasons of the market. From the observation that the skewness gradually drops if we extend the measurement window of returns, we can conclude that the asymmetry of returns is not caused by the possibility that there is more or stronger good news than bad news in the Chinese markets. It is the asymmetry of response of the market to the news which leads to the asymmetry of returns.

\begin{figure}
\centering 
\includegraphics[width =0.7\textwidth]{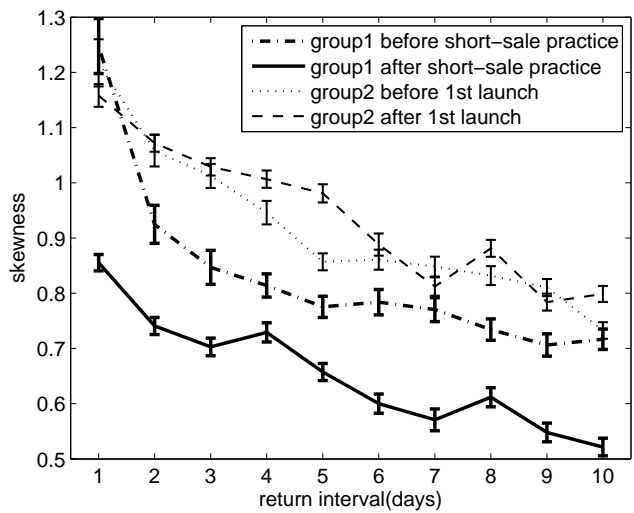}
%\caption{The change of skewness due to short sale}
\caption {The average skewness and its sample standard deviation (divided by the square root of sample size minus 1) as the function of return intervals. Group 1 refers to the stocks which are on the designated list at least once as of March 31, 2014, Group 2 refers to the stocks which are not. The bottom line presents the skewness of Group 1 after short sales are practice. The second line from the bottom presents the skewness of normalized excess return of Group 1 before they can be shorted. Since stocks in Group 2 do not have an initiating date for short sale, the top two lines stand for the skewness of their returns before and after the first launch of the security lending policy on March 31, 2010. Error bars denotes one standard deviation range of skewness.}
\label{fig:skewness}
\end{figure}

\begin{figure}
\centering 
\includegraphics[width =0.7\textwidth]{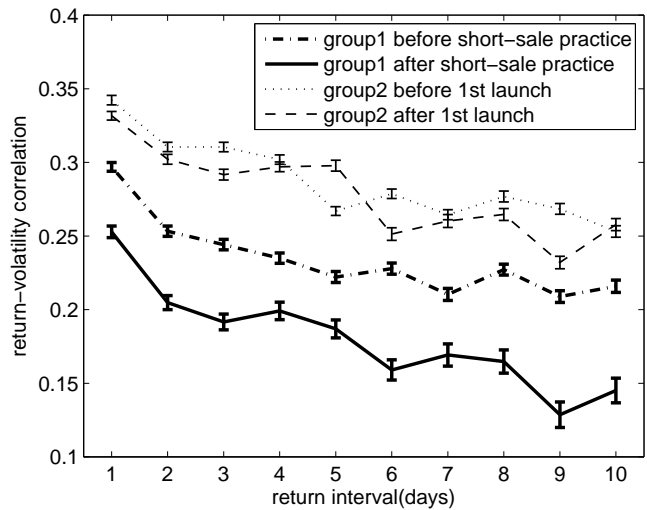}
%\caption{The change of volatility-return correlation due to short sale}
\caption{The average correlation of return-volatility correlation (excess returns) and its sample standard deviation (divided by the square root of sample size minus 1) as the function of return intervals. Group 1 refers to the stocks which are on the designated list at least once as of March 31, 2014, Group 2 refers to the stocks which are not. The bottom line presents the correlation of Group 1 after short sales are practiced. The second line from the bottom presents the correlation of normalized excess return of Group 1 before they can actually be shorted. Since stocks in Group 2 do not have an initiating date for short sale, the top two lines stand for the correlation of their returns before and after the first launch of the security lending policy on March 31, 2010. Error bars denotes one standard deviation range of correlation.}
\label{fig:correlation}
\end{figure}

Similar to the case for skewness, in Figure \ref{fig:correlation} we report a significant decrease in the return-volatility correlation for Group 1 after short sales are practiced. In contrast, the two intertwined lines at the top show that the volatility-return correlation of Group 2 does not change much after the implementation of the security lending policy since they are excluded from the designated list.
%This result shows that the positive skewness is linked to the positive volatility-return distribution of stocks, and short sale restrictions are also a cause for that. Just like skewness, volatility-return correlation also decreases with the increase of return interval, due to the gradual assimilation of information.  

\subsection{Panel Regression}
We have described the decrease of skewness and return-volatility correlation as a result of the security lending policy in China in Figure \ref{fig:skewness} and \ref{fig:correlation}. We depict the dramatic change of the considered statistics before and after short sale practice for stocks allowed for short sales in group 1, in contrast to the unnoticed change for stocks not eligible for short sales in group 2. The possibility still exists that this phenomenon might be caused by the mere passage of time or other endogenous factors which we have not been able to identify. Therefore, we formalize the problem in the framework of panel regression for more rigorous analysis. 

We use all the statistics as the dependent variables in Table \ref{tab:des}, namely the skewness of three different measures, leverage coefficient $\xi_1$ in the EGARCH model, and two return-volatility correlations. Explanatory variables include turnover rate (which is the ratio of daily volume to shares outstanding) to proxy for the difference of opinions, return and volatility for each stock, a dummy variable $v$ to represent group 1 or group 2 and a dummy variable $u$ to represent short sale practiced. 
The other two sets of dummy variables account for the fixed effects of industry and time. There are 13 industry categories in total according to the national 2001 criteria of stock classification of the CSRC (China Securities Regulatory Commission). The number of time windows depends on how we prepare our samples.

All dependent variables and explanatory variables are estimated in non-overlapping time windows spanning from January 1, 2006 to March 31, 2014. 
As a robustness check, we use two time windows to prepare the samples (i.e., quarterly and half-year time windows). If the samples are calculated for every quarter, there are 33 samples for each stock, denoting 33 quarters from January 1, 2006 to March 31, 2014. If we prepare the samples for every half-year, then there are 17 samples.

The regression equation is given as, 
\begin{equation}
y_{it} = \alpha_I + \gamma_t + \beta_T T_{it} + \beta_r r_{it} + \beta_\sigma \sigma_{it} + \beta_v v_i + \beta_u u_{it}
\end{equation}
where $y$ is one of the 5 dependent variables, $\alpha_I$ and $\gamma_t$ is the fixed effects of industry and time, respectively. $T_{it}$, $r_{it}$ and $\sigma_{it}$ represent the turnover rate, return and volatility of stock $i$ in time window $t$, respectively. Dummy variable $v_i=1$ means stock $i$ is in group 1, and $u_{it}=1$ means stock $i$ can be sold short in time window $t$. 
As we have noticed in Table \ref{tab:events}, the designated list is frequently adjusted. Some stocks are later included again after they have been removed.
Remember that group 1 refers to stocks that are selected for short sales practice at least once as of March 31, 2014. There are stocks in group 1 which are not practiced for short sales yet in time window $t$. But for those stocks in group 2 ($v_i = 0$), we always have $u_{it} = 0$.
To determine whether a stock in group 1 is practiced for short sale or not, we count the actual number of days for each stock in each time window.
If a stock is eligible for short selling for more than half of the days in the given time window, we treat it as a sample with short sale practiced.

The regression results are reported in Table \ref{tab:quarter} and Table \ref{tab:half-year} for quarterly and half-year sampling periods, respectively. The results are similar despite the sampling windows. As shown in the Tables, all dependent variables except for the skewness of raw returns decreased significantly at a 1\% level after short sale is practiced. These results give the direct evidence to support our hypothesis that the short sale constraint is responsible for the statistical anomalies we have observed in the Chinese markets. The skewness of raw returns also decreases but to a lesser degree, which could explain why we have conflicting results in \cite{chang2014short} and \cite{zhao2014short} as they report the skewness of raw returns.

Turnover rate (the proxy for difference of opinion) positively contributes to the dependent variables in accordance with the proposition made by \cite{hong2003differences}: that difference of opinion results in higher skewness. It is interesting to note that the contribution of return to dependent variables is positive (i.e., we expect to get higher skewness, stronger anti-leverage effect and higher risk-return correlation when the price of a stock is pumped up). This could be explained by the price chasing behaviors of investors in China. When the short sales are constrained, investors would rush for price rising to make profits since they can only buy low and sell high (they can not sell high and buy low). The results are inconclusive for volatility. %The coefficients have the predicted sign and are highly significnat.

% This figure With no regarding to the time nor the group of stocks, we simply mix up the normalized innovations of all stocks from 2006 to 2014 and get the figure of distribution in the left part of Figure 1, from which the asymmetry can be clearly seen. A larger number of stocks are presenting positive innovations of a larger magnitude. In the right part of Figure 1, we calculate the cumulative probability of the distribution, again there's a perspicuous discrepancy between the real values and the presumed values where the excess return is symmetric.The departure of the distribution in reality from the distribution it should be if it is symmetric leads to the conclusion that the excess return is asymmetric.

\begin{landscape}
\begin{table}
\centering
%  \caption{Quarterly panel data regression}
\caption {The table reports the regression conducted on quarterly samples if there are more than 40 observations in a quarter. Dependent variables are skewness of raw returns, excess returns, normalized excess return, leverage coefficient $\xi_1$ in EGARCH model, return-volatility correlation calculated for raw returns and excess returns, respectively. Explanatory variables include, average daily turnover in a quarter, cumulative return and standard deviation of daily return in a quarter, a dummy variable for designated list, and a dummy variable for short sale practice. All regressions include dummies (unreported) for each quarter and industry.}
   \begin{tabular}{lcccccl}
    \hline
    &\multicolumn{3}{c}{skewness} & leverage & \multicolumn{2}{c}{return-volatility correlation} \\
    \cline{2-4}
    \cline{6-7}
	& raw & excess & normalized & coefficient & raw & excess \\
      %\ & skewRaw & skewRelative & skewNormalized & Leverage & & Volatility-return Corr \\
    \hline
    turnover rate  & 0.0076*** & 0.031*** & 0.037*** & 0.014*** & 0.022*** & 0.021***\\
    \         & (4.48) & (13.12) & (13.44) & (12.71) & (27.22) & (22.02) \\
    \hline
    return   &  0.010***  & 0.021*** & 0.041*** & 0.006*** & 0.009*** & 0.03***\\
    \         & (5.91) & (8.50)  & (14.95) & (6.10) & (11.81) & (31.22) \\
    \hline
    volatility   & 0.78**  & -0.10** & -0.078*** & 0.03 & 0.77*** & -0.29\\
    \         & (2.22) & (-2.03) & (-3.20) & (0.13) & (4.68) & (-1.49) \\
    \hline
    designated list         & 0.070***  & -0.053*** & -0.078***  & -0.026*** & 0.060*** & -0.0062* \\
    \         				& (11.54) & (-6.25) & (-8.08) & (-7.01) & (21.65) & (-1.88) \\
    \hline
    short sale practiced         & -0.017 & -0.056*** & -0.081*** & -0.026*** & -0.026*** & -0.014**\\
    \         						& (-1.55) & (-3.60) & (-4.60) & (-3.90) & (-5.29) & (-2.37) \\
    \hline
    Obj.      & 38142 & 38142 & 38142 & 38142 & 38142 & 38142\\
    Stocks    & 1357 & 1357  & 1357  & 1357 & 1357 & 1357\\
    R-square adjusted & 0.071 & 0.044 & 0.049  & 0.027 & 0.30 & 0.091\\
    F-statistics & 59.12 & 35.53 & 39.98 & 19.77 & 307.59 & 77.38\\
    Time fixed effect & YES & YES & YES & YES & YES & YES\\
    Industry fixed effect & YES & YES & YES & YES & YES & YES\\
    \hline
  \end{tabular}
\small{\emph{Note:Triple asterisks(***)denote significance at 1\% critical level, double asterisks(**) at 5\% level, and single asterisks(*) at 10\% level respectively.}}

\label{tab:quarter}
\end{table}
\end{landscape}

\begin{landscape}
\begin{table}
  \centering
%  \caption{Half-year panel data regression}
  \caption {The table reports the regression conducted on half year samples if there are more than 40 observations in half year. Dependent variables are skewness of raw returns, excess returns, normalized excess return, leverage coefficient $\xi_1$ in EGARCH model, return-volatility correlation calculated for raw returns and excess returns, respectively. Explanatory variables include, average daily turnover in half a year, cumulative return and standard deviation of daily return in half a year, a dummy variable for designated list, and a dummy variable for short sale practice. All regressions include dummies (unreported) for each half year and industry.}  
  \begin{tabular}{lcccccl}
    \hline
    &\multicolumn{3}{c}{skewness} & leverage & \multicolumn{2}{c}{return-volatility correlation} \\
    \cline{2-4}
    \cline{6-7}
	& raw & excess & normalized & coefficient & raw & excess \\
      %\ & skewRaw & skewExcess & skewNormalized & Leverage & Volality-return Corr  \\
    \hline
    turnover rate & -0.0073*** & 0.022*** & 0.027*** & 0.014*** & 0.011*** & 0.016***\\
    \         & (-2.60) & (5.67) & (6.26) &  (10.32) & (10.49) & (13.15)\\
    \hline
    return   & 0.0055*** & 0.020*** & 0.036*** & 0.0058*** &  0.0055*** & 0.022***\\
    \         & (2.98) & (7.84) & (12.92) & (7.042) & (8.72) & (27.62)\\
    \hline
    volatility   & 1.53*** & -0.96  & -1.44* & -0.15 & 1.51*** & 0.0093 \\
    \         & (2.88) & (-1.34) & (-1.81) & (-0.61) & (7.98) & (0.041) \\
    \hline
    designated list         & 0.064*** & -0.065*** & -0.084*** & -0.021*** & 0.053*** & -0.0058 \\
    \         				& (6.88) & (-5.22) & (-6.00) & (-4.97) & (16.40) & (-1.49) \\
    \hline
    short sale practiced         & -0.029* & -0.08*** & -0.12*** & -0.025*** & -0.036*** & -0.023*** \\
    \         						& (-1.82) & (-3.72) & (-4.99) & (-3.41) & (-6.46) & (-3.39) \\
    \hline
    Obj.      & 19372 & 19372 & 19372 & 19372 & 19372 & 19372 \\
    No. of Stocks    & 1357 & 1357 & 1357 & 1357 & 1357 & 1357 \\
    R-square adjusted  & 0.061 & 0.044 & 0.051 & 0.025 & 0.30 & 0.11  \\
    F-statistics & 38.78 & 27.27 & 31.83 & 14.62 & 249.39 & 70.42  \\
    Time fixed effect & YES & YES & YES & YES & YES & YES \\
    Industry fixed effect & YES & YES & YES & YES & YES & YES \\
    \hline
  \end{tabular}
\small{\emph{Note:Triple asterisks(***)denote significance at 1\% critical level, double asterisks(**) at 5\% level, and single asterisks(*) at 10\% level respectively.}}
  \label{tab:half-year}
\end{table}
\end{landscape}

\section{Discussion and conclusion}
Some statistical anomalies, including positive skewness, anti-leverage effect and reverse volatility asymmetry (positive return-volatility correlation) have been documented on daily firm-level data on the Chinese stock market. We show empirically that short sale constraints are responsible for these anomalies. 

We first examine the skewness of daily firm-level returns in the Chinese stock market. The skewness of raw returns, excess returns and normalized excess returns are all positive in that the mean and median of skewness are statistically significantly positive. The skewness of the three returns captures different statistical aspects of the market. The skewness of the excess returns is particularly significantly positive in both its mean and median after the market part is removed. The effect of time-varying volatility and leverage (or anti-leverage) effects are removed after we normalize the excess returns using the conditional volatility generated by the EGARCH model. The skewness changes very little after the excess returns are normalized, suggesting that the time-varying volatility and leverage (or anti-leverage) effect are not the cause of positive skewness of excess returns. 

As for the risk-return relationship, the positive mean and median of the leverage coefficient of the EGARCH model indicates that we have an anti-leverage effect in the Chinese markets. The leverage coefficient measures the effect of lagged returns on future volatility. Anti-leverage effect means that positive returns induce higher volatility than negative returns. Contemporaneous return-volatility correlation, however, captures the simultaneous co-movement of returns and volatility at a daily level although the dynamic nature of the interaction can be complex at the intraday level. If the contemporaneous correlation of return and volatility is larger than zero, we would expect to get larger returns when the returns are positive than when the returns are negative and therefore, positive skewness is a direct consequence of the positive return-volatility correlation. 

The two theories (i.e., the leverage effect and volatility feedback effect) developed to explain the leverage effect and volatility asymmetry in the U.S. markets apparently do not apply to the Chinese markets. In this study, we empirically verify that the positive skewness is caused by the short sale constraints as an empirical support to the model developed by Xu(2007). We also provide evidence that the anti-leverage effect and positive return-volatility correlation are also caused by short sale constraints. 

Starting from March, 2010, China adopts the margin finance and security lending policy. The policy provides for a designated list of stocks for short sale practice.
The adoption of the security lending policy takes place over time and it is still in process. This affords us sufficient time space from which to collect data to examine the effects of the policy. We divide the Chinese stocks into two groups, namely Group 1 and Group 2. Group 1 includes stocks that are selected for short sales practice at least once as of March 31, 2014; Group 2 includes stocks that are never eligible for short sales. Graphically, we show a significant reduction of the skewness and return-volatility correlation for the stocks in Group 1 after the short sales are practiced. In addition, we show that the skewness and return-volatility correlation also gradually drop if we extend the measurement window of returns from 1 day to 10 days. This suggests that the asymmetry of returns is not caused by the possibility that in the Chinese markets we are populated with more good news or the intensity of good news is stronger than bad news.
%have more or stronger good news than bad news. 
It is, rather, the asymmetric response of the market to the news which causes the asymmetry of returns. 

We then conduct panel regression using daily return data of stocks allowed for short sale after they are added to a designated list and of other stocks never added to the list. Our results show that (i) the lifting of short sale constraints largely not only decreases the skewness of the raw returns, excess returns and normalized excess returns, but also decreases the anti-leverage effect and reverse volatility asymmetry; (ii) the skewness, anti-leverage effect, and reverse volatility asymmetry are all positively correlated with turnover rate (proxy of difference of opinion), and with contemporaneous returns. 

Positively skewed securities are like lotteries. \cite{stockaslottery} show that investors are skewness seekers if they obey prospect theory (\cite{RePEc:cla:levarc:7656}). Chinese investors would rather pay high prices to hold securities with high skewness and treat those stocks as lotteries. It has been shown by many researchers (see e.g., \cite{chineselottery}) that there exists obvious lottery preference in the Chinese markets. The significant positive skewness of returns provides investors with the motive to chase for lottery-like stocks. Investors are encouraged to bet on the continuation of price rising in the Chinese markets where the stock price returns skew more to the upside when short sales are constrained. %Our research sheds a light on a possible cause of this phenomenon. 

%\section{acknowledgments}
%The partial financial support from the Fundamental Research Funds for the Central Universities under grant number skyb201403, and the Start up Funds from Sichuan University under grant number yj201322 is gratefully acknowledged.

\section{References}
%\begin{footnotesize}
\bibliographystyle{elsarticle-harv}
\bibliography{firstbib}
%\end{footnotesize}

\end{document}